\documentstyle[11pt,aaspp4,psfig,flushrt]{article}

\def\simg{\mathrel{%
      \rlap{\raise 0.511ex \hbox{$>$}}{\lower 0.511ex \hbox{$\sim$}}}}
\def\siml{\mathrel{%
      \rlap{\raise 0.511ex \hbox{$<$}}{\lower 0.511ex \hbox{$\sim$}}}}
\def\gta{\simg} \def\lta{\siml}
\def\etal{et al$.$ } \def\eg{e$.$g$.$ } \def\ie{i$.$e$.$ }
\def\Meszaros{M\'esz\'aros }
\def\E{{\cal E}}
\def\B4{\epsilon_{B_{_{-4}}}}
\def\eB{\epsilon_B}
\def\eBr{\epsilon_{Br}}

\def\eBf{\epsilon_{Bf}}
\def\ni{\noindent}
\def\o{\over}
\def\qquad{\quad\quad}
\def\qqqquad{\quad\quad\quad\quad}

\begin{document}

\title{A Unified Treatment of the Gamma-Ray Burst 021211 and Its 
   Afterglow Emissions} 

\author{Pawan Kumar \& Alin Panaitescu}
\affil{Astronomy Department, University of Texas, Austin, TX 78731}

\begin{abstract}
The Gamma-Ray Burst (GRB) 021211 detected by the High Energy Transient 
Explorer (HETE) II had a simple light-curve in the x-ray and gamma-ray 
energy bands containing one peak and little temporal fluctuation
other than the expected Poisson variation. Such a burst offers the 
best chance for a unified understanding of the gamma-ray burst and 
afterglow emissions. We provide a detailed modeling of the observed 
radiation from GRB 021211 both during the burst and the afterglow phase. 
The consistency between early optical emission (prior to 11 minutes), 
which presumably comes from reverse shock heating of the ejecta, and 
late afterglow emission from forward shock (later than 11 minutes) 
requires the energy density in the magnetic field in the ejecta, 
expressed as fraction of the equipartition value or $\epsilon_B$, to 
be larger than the forward shock at 11 minutes by a factor of about 10$^3$. 
We find that the only consistent model for the gamma-ray emission in
GRB 021211 is the synchrotron radiation in the forward shock; to explain
the peak flux during the GRB requires $\epsilon_B$
in forward shock at deceleration to be larger than the value at 11 minutes
by a factor of about 10$^2$.

These results suggest that the magnetic field in the reverse shock and early
forward shock is most likely frozen-in-field from the explosion, and 
therefore a large fraction of the energy in the explosion was initially 
stored in magnetic field.
We can rule out the possibility that the ejecta from the burst for GRB 021211 
contained more than 10 electron-positron pairs per proton.

\end{abstract}


\section{Introduction}

Considerable progress has been made in the last few years toward an 
understanding of the nature of the enigmatic Gamma-Ray Bursts. 
Much of this progress has resulted from the
observation and analysis of afterglow emission - the radiation we receive 
after the high energy gamma-ray photons ceases  - which is firmly 
established to be synchrotron radiation from a relativistic, external shock
(\eg Wijers, Rees, and \Meszaros 1997).
The nature of the explosion and the process that generates gamma-ray
photons continue to be debated, although it is widely accepted 
that these explosions involve a stellar mass object. The detection of
narrow emission lines (\eg Greiner \etal 2003) and the emergence of a 
spectrum similar to that of SN 1998bw (Matheson \etal 2003) in the afterglow 
of GRB 030329 indicate that at least some GRBs are produced when a 
massive star undergoes collapse at the end of its nuclear burning life. 

Further progress toward understanding the GRB explosion 
requires afterglow observations at times closer to the burst and a 
simultaneous modeling of both the afterglow and gamma-ray emissions. 
In this way, one can explore distance scales of $\sim 10^{16}$cm from the
center of explosion, \ie an order of magnitude smaller than that probed
by afterglow emissions at half a day or later. Such a treatment is more
likely to succeed in those cases where the prompt (burst) emission arises 
from the same region as the delayed (afterglow) emission, \ie from
an external shock. The simple FRED-like (fast rise, exponential decay)
light-curves seen in about 10\% of bursts represents the type expected 
from an external shock (\Meszaros \& Rees 1993), while short variability 
timescale bursts with complicated light curves are usually attributed to 
internal shocks in an unsteady outflow (Rees \& \Meszaros 1994; see Piran 
1999 \& \Meszaros 2002 for recent reviews). 

This paper is an attempt to explain with the same process -- synchrotron and
inverse Compton emission from an external shock -- the burst and afterglow 
emission of GRB 021211 detected by the HETE II (Crew \etal 2003), a burst which
had a simple, FRED-like morphology and whose afterglow has been followed 
starting from 60 seconds until 10 days after the burst. 
In \S\ref{obs} we summarize the observations of GRB/afterglow 021211. 
In \S\ref{eqs}--\S5 we present the formalism for calculating the synchrotron 
and inverse Compton emissions from both the forward and reverse shocks, and in
\S\ref{s0} and \S\ref{s2} we assess the ability of the synchrotron, self-Compton 
model with a uniform and an $r^{-2}$ stratified medium to accommodate the 
properties of the GRB 021211 and its afterglow.

\section{Summary of Observations for GRB 021211}
\label{obs}

 At a redshift $z=1.0$, GRB 021211 had a duration of $\gta 2.3$s
in the 30-400 kev energy band and a fluence of $\sim 2\times 10^{-6}$ erg
(Crew \etal 2003); the duration in the 10-25 kev band was $\sim 4$s.
If both the burst and the afterglow for GRB 021211 arise from some combination 
of reverse and forward external shocks, then the deceleration time $t_d$ 
is close to the time when the GRB light-curve peaks in 30-400 kev band, 
\ie about 2 seconds. 
The average flux during the first 2.3s was 4 mJy in the 7-30 kev band,
3 mJy in the 50-100 kev band, and 0.5 mJy in at 100-300 kev, while the peak
of the $\nu F_\nu$ spectrum was at $47^{+9}_{-7}$ kev. The isotropic equivalent 
of the energy released in the 10-400 kev emission is $\sim 10^{52}$ erg.

 At 90s after the burst, the R-band magnitude of the afterglow 021211 was 
14.06 (Wozniak \etal 2002), corresponding to a flux of 7.2 mJy. The optical 
flux decayed as $t^{-1.82 \pm 0.02}$ for the first 10 minutes, after which it 
flattened to a $t^{-0.82 \pm 0.11}$ fall-off (Li \etal 2003), reaching
magnitude 25 at 7 days (Fruchter \etal 2002). The steeper decay seen during 
the first 10.8 minutes suggests that the optical emission is dominated by 
the reverse shock energizing the GRB ejecta, while the shallower, later 
time decay is attributed to the forward shock that sweeps-up
the ambient medium. The R-band flux at 11 minutes, when the two 
contributions are equal, is 0.39 mJy, therefore the forward shock 
optical flux at 11 minutes is 0.19 mJy. Fox \etal (2003) report a 
3-$\sigma$ upper limit of 110 $\mu$Jy on the radio (8.5 GHz) flux 
at 0.1 days, and an upper limit of 35 $\mu$Jy during 9--25 days.

Finally, Milagro has reported an upper limit of $4\times 10^{-6}$erg cm$^{-2}$
on the 0.2-20 Tev fluence over the burst duration reported by the HETE WXM
 (McEnery \etal 2002).

\section{Shock dynamics \& deceleration time}
\label{eqs}

Consider an explosion where the isotropic equivalent of energy release is $\E$ 
and the initial Lorentz factor (LF) of cold baryonic material carrying this 
energy is $\Gamma_0$. Before the ejecta are significantly decelerated, the 
thermal LF $\gamma_{p,f}$ of the protons in the forward shock (FS), equal 
to the the bulk LF $\Gamma_d$ of the swept-up medium, is (\eg Piran 1999)
\begin{equation}
 \gamma_{p,f} = \Gamma_d \approx (\Gamma_0/2)^{1/2} (n_{ej}/n)^{1/4},
\label{gmpf}
\end{equation}
where $n_{ej}$ is the comoving particle density of the ejecta, 
and $n = (A/m_p) r^{-s}$ is the radial profile of the external 
particle density ($s=0$ for a homogeneous medium, $s=2$ for a 
pre-ejected wind). The above result holds for $n_{ej}/n\lta \Gamma_0^2$, 
otherwise $\gamma_{p,f}\approx \Gamma_0$. 

Taking into account that the laboratory frame energy per FS-heated 
proton is $\Gamma_d \gamma_{p,f}=\gamma_{p,f}^2$, the deceleration radius 
$R_d$ at which the energy of the swept-up medium is half the initial energy 
of the ejecta is 
\begin{equation}
{4\pi\over 3-s} R_{d}^{3-s} A c^2 \Gamma_0\left({n_{ej}\over n} \right)^{1/2}
     = \E.
\label{rda}
\end{equation}
As long as the distribution of LF of the ejecta is not too narrow, and the 
duration of the central explosion is less than $R/(c\Gamma_0^2)$, the 
comoving width of the material ejected in the explosion is proportional 
to $R/\Gamma_0$. Let us parametrize the comoving thickness of the ejecta 
as $\eta R/\Gamma_0$. Therefore, the comoving density of the ejecta is
\begin{equation}
 n_{ej}=\frac{\E}{4\pi R^3\eta m_p c^2}, 
\label{nej}
\end{equation}
which substituted in equation (\ref{rda}) leads to
\begin{equation}
 R_d =c\left[ { (3-s)^2 \eta \E\o 4\pi A c^2 \Gamma_0^2 }\right]^{1\o3-s}.
\label{Rda}
\end{equation}
This result is identical to that obtained for the Blandford-McKee 
self-similar solution extrapolated back to $R_d$ if we set $\eta=17/18$ 
for $s=0$ and $\eta=9/2$ for $s=2$. 

 From equations (\ref{nej}) and (\ref{Rda}), the comoving density of the ejecta at $R_d$ is 
\begin{equation}
{n_{ej}\over n} (R_d) = {\Gamma_0^2\over (3-s)^2\eta^2} ,
\label{nejRa}
\end{equation}
for $\eta \gta 1$.
Substituting this into equation (\ref{gmpf}), the LF of the shocked ISM 
at $R_d$ is
\begin{equation}
\Gamma_d = {\Gamma_0\over \left[ 2(3-s)\eta\right]^{1/2}}.
\label{gammada}
\end{equation}
 From equations (\ref{Rda}) and (\ref{gammada}), the observer-frame 
deceleration timescale is
\begin{equation}
 t_d = (1+z) f_\eta {R_d \over c \Gamma_d^2} = {(1+z) f_\eta \over c\,
   \Gamma_0^2}\left[ {(3-s)^{5-s}\eta^{4-s}\E\over 4\pi Ac^2\Gamma_0^2} 
   \right]^{1\o3-s},
\label{tda}
\end{equation}
where $f_\eta$ is a correction factor that takes into account the 
difference between the arrival time $(1+z) R_d /(2 c \Gamma_d^2)$ of 
photons emitted from the contact 
discontinuity\footnotemark \footnotetext{The usual factor 2 in the 
denominator, corresponding to photons moving along the direction 
observer--center of explosion, is compensated by that most emission 
arises from the gas moving at an angle $\Gamma_d^{-1}$ relative to that 
direction.} and that from where most of the GRB emission arises\footnotemark. 
\footnotetext{For instance, if the burst is FS synchrotron emission from 
higher energy electrons in a fast cooling regime, then the $\gamma$-ray 
emission arises from the shocked gas immediately behind the FS and 
$f_\eta = 1/4$. At the other extreme, when the burst arises from fast 
cooling electrons located immediately behind the reverse shock, it can 
be shown that $f_\eta \simeq (1/2) \sqrt{\eta/(3-s)}$ for $\eta \gg 1$.} 
The lab frame speed of the reverse shock relative to the back-end of the shell,
shown in figure 1, is $V_{rs,lab}/c \approx 1.4 (3-s)\eta/\Gamma_0^2$. Thus,
the time it takes for the RS to cross the shell (in lab frame) is, 
$\eta R_d/(\Gamma_0^2 V_{rs,lab})\approx [1.4(3-s)]^{-1} R_d/c$. And so the RS
crossing time is same as the deceleration time to within a factor of order 
unity.

For $t>t_d$ the LF decreases as 
\begin{equation}
\Gamma(t)=\Gamma_d \left({t\over t_d}\right)^{-{3-s\over 8-2s}}.
\label{gamFS}
\end{equation}

\section{Forward Shock}

The comoving density behind the forward shock (FS) is, 
$\rho=4\rho_0\Gamma$, and the thermal energy density is $u=4\rho_0 
c^2\Gamma^2$; where $\rho_0=A r^{-2}$ is the density of the medium 
just ahead of the shock, and $\gamma$ is the bulk LF of shocked
fluid given by equation (\ref{gamFS}). A fraction $\epsilon_e$ of 
the thermal energy of the shock-heated circumburst medium is taken 
up by electrons. Electrons with thermal LF greater than $\gamma_i$
are assumed to have a powerlaw distribution with index $p$, \ie
$dN_e/d\gamma\propto \gamma^{-p}$ for $\gamma>\gamma_i$, where $\gamma_i=
\epsilon'_e (m_p/m_e) \gamma_p$, is the minimum thermal LF of
electrons; $\epsilon_e' = [(p-2)/(p-1)]\epsilon_e$ for $p > 2$ and 
$\gamma_p$ is the proton thermal LF. The 
energy density in magnetic field is assumed to be $\epsilon_{B_f} u$, 
and therefore the magnetic field is $B = 4\Gamma c(2\pi\epsilon_{B_f} 
A r^{-2})^{1/2}$. 

The FS synchrotron {\bf injection} frequency, $\nu_{if}$ and the flux at 
the peak of the $F_\nu$ spectrum, are 
\begin{equation}
\nu_{if}(t)={q B\gamma_i^2\Gamma\over 2\pi m_e c(1+z)} = {4q\epsilon_e'{^2}
    \epsilon_{B_f}^{1\o 2}m_p^2 A^{1\o 2}\Gamma_d^{4-s}\over\sqrt{2\pi}\,
      m_e^3 (1+z)}\left[{4(4-s)c t_d\over 1+z}\right]^{-s/2}
     \left({t\over t_d}\right)^{-3/2},
\label{num}
\end{equation} 
\begin{equation}
f_{\nu_p,f}(t) = {N_e \epsilon'_{\nu'_p}\Gamma\over d_L^{'^2} }=
    {4 (6\pi)^{1/2} q^3 A^{3\o 2}\epsilon_{B_f}^{1\o 2}\Gamma_d^{8-3s}
      \over m_p m_e c (3-s) {d_L'}^2} \left[ {4(4-s)c t_d\over 1+z}
      \right]^{3(2-s)\o 2} \left( {t\over t_d}\right)^{-{s\over 8-2s}},
\label{fnu}
\end{equation}
where $q$ \& $m_e$ are electron charge and mass, $m_p$ is proton mass,
and $d_L' = D_L/\sqrt{1+z}$, $D_L$ being the luminosity distance, $N_e =
A R^{3-s}/(3-s)$ is the number of electrons per unit solid angle behind
the shock, and $\epsilon'_{\nu'_p}=3^{1/2} q^3 B/m_e c^2$ is power per unit 
frequency per electron, in comoving frame, at the peak of the synchrotron
spectrum. 

The synchrotron {\bf injection} frequency for the cases of $s=0$ \& 2 are
written out explicitly for ease of application later on
\begin{equation}
  \nu_{if}(t) =
     {\epsilon_e'}^{2}\epsilon_{B_{f_{-4}}}^{1\o 2}(t/t_d)^{-{3\o 2}}
     \times\Biggl\{    {\parbox{7.7cm}{\begin{eqnarray}
            3.7\times10^{18} n_0^{1/2}\Gamma_{d_2}^4 (1+z)^{-1} \;{\rm Hz}
                       \quad\quad\quad  {\rm s=0} \nonumber\\
            1.7\times10^{21} A_*^{1/2}\Gamma_{d_2}^2 t_d^{-1} \;{\rm Hz}
                \;\quad\quad\quad\quad\quad{\rm s=2} \nonumber
       \end{eqnarray}}  }
        \Biggr.
\label{numa}
\end{equation}
where $A_* = A/5$x$10^{11}$ g cm$^{-1}$, and an integer subscript $n$ on a 
variable $X$, $X_n$, means $X/10^n$.
The flux at the peak of the synchrotron spectrum for $s=0$ \& 2 is
\begin{equation}
f_{\nu_p,f}(t) = \epsilon_{B_{f_{-4}}}^{1\o 2}\Gamma_{d_2}^2 \times\Biggl\{ 
     {\parbox{6.6cm}{ \begin{eqnarray}
        5.5\times10^{-7} n_0^{3/2}\Gamma_{d_2}^6 t_d^3 \;{\rm mJy}
          \quad\quad\quad{\rm s=0}\nonumber \\
           1.2\times10^{3} A_*^{3/2} (t/t_d)^{-1/2} \;{\rm mJy}
               \;\quad{\rm s=2} \nonumber
      \end{eqnarray}  } }
       \Biggr.
\label{fnua}
\end{equation}
In the derivation of the above equation we have set $z=1$ which corresponds 
to the redshift of GRB 021211.

The FS synchrotron self-absorption frequency ($\nu_{Af}$), obtained by equating
the intensity at $\nu_{Af}$ to $2 m_e\gamma_{if} \nu_{Af}^2$, is given by
\begin{equation}
\nu_{Af}^2 \left( {\nu_{if} \over \nu_{Af}}\right)^\alpha = {(6\pi\eB)^{1/2} q^3
    A^{3\o 2} \Gamma_d^{4-3s} (4t_d)^{(2-3s)/2}\over 8(3-s)m_e m_p^2
    \epsilon'_e (1+z)^{(6-3s)/2} c^{3s/2}} \left({t\over t_d}\right)^{-{s+4
    \over 8-2s}},
\label{nuAFS}
\end{equation}
where $\alpha$ depends on the relative location of $\nu_{Af}$ with respect with 
$\nu_{if}$ and the cooling frequency $\nu_{cf}$; for $\nu_{cf}>\nu_{if}
>\nu_{Af}$, $\alpha=1/3$, and $\alpha=p/2$ if $\nu_{if}<\nu_{Af}<\nu_{cf}$.

\subsection{Application to GRB 021211 late time optical observation}

We make use of the optical R-band flux at late time, $t\ge11$min, to
provide constraints on the density, $\epsilon_{Bf}$ \& $\Gamma_d$ for
$s=0$ and $s=2$ cases separately.

\subsubsection{Parameters for a uniform ISM model}

Using equation (\ref{tda}) the bulk LF at deceleration time, for $s=0$,
is found to be
\begin{equation}
\Gamma_{d_2} = 3.8 \left({\E_{52}\over n_0}\right)^{1/8} \left[{f_\eta(1+z)
    \over t_d}\right]^{3/8}, 
\label{tda0}
\end{equation}
where $n_0$ is density of the uniform ISM, $f_\eta=\max\{1,2\sqrt{\eta}/3\}$
(see footnote 2) and an integer subscript on a variable $X_n$ means $X/10^n$.

The observed R-band flux at 11 min for GRB 021211 is 0.4 mJy. According
to the fit presented in Li \etal (2003) the contributions from the reverse
and forward shocks to the observed R-band flux are equal at this time.
Therefore, the FS peak flux at 11 min is greater
than 0.2 mJy; we will consider the peak flux to be $0.2\, A_f$ mJy, with 
$A_f>1$. It should be noted that for $s=0$ the peak flux is time
independent. Substituting this into equation (\ref{fnua}), and making use of 
equation (\ref{tda0}) to eliminate $\Gamma_d$ we find
\begin{equation}
  n_0^{1\o 2}\E_{52}\epsilon_{B_{f_{-4}}}^{1\o 2} = A_f f_\eta^{-3}.
\label{cons1}
\end{equation}
The R-band lightcurve is observed to be monotonically declining from the
earliest time (90s), and from 11 min to 10 days the decline is a simple
powerlaw with index $0.82\pm0.11$. Thus, the frequency of the peak of the 
spectrum at 11 min is expected to be less than the R-band frequency 
of 4.7x$10^{14}$ Hz or 1.95 ev. Let us assume that the peak frequency at 
11 min is a factor $A_\nu$
smaller than the R-band frequency. Substituting this into equation (\ref{numa})
and making use of equation (\ref{tda0}) we find
\begin{equation}
\E_{52}^{1/2} \epsilon_{B_{f_{-4}}}^{1/2}\epsilon_e'{^2}f_\eta^{3/2} = 
   8\times10^{-3} A_\nu^{-1} \quad{\rm or} \quad 
    \epsilon_{B_f}(t=11min) = 6.4\times10^{-9} A_\nu^{-2} \E_{52}^{-1} 
        \epsilon_e'{^{-4}} f_\eta^{-3}.
\label{cons2}
\end{equation}
The synchrotron peak frequency as a function of time is given by
\begin{equation}
\nu_{if} = 37.3 \,t^{-3/2} A_\nu^{-1} \; {\rm kev},
\label{num0}
\end{equation}
Combining equations (\ref{tda0}), (\ref{cons1}) and (\ref{cons2}) we find
\begin{equation}
\Gamma_{d_2}^2 \approx 2.2\, (A_\nu A_f)^{-1/2}f_\eta^{3/2} \E_{52}^{1/2}
  \epsilon_e'^{-1} t_d^{-3/4}.
\label{gamd20}
\end{equation}
Substituting this back into equation (\ref{tda0}) we obtain
\begin{equation}
n_0 = 1.5\times10^4\, (A_\nu A_f)^{2} \E_{52}^{-1}f_\eta^{-3}\epsilon_e'{^4}
     \;\; {\rm cm}^{-3}.
\label{n0}
\end{equation}
\ni Note that $A_f$ and $A_\nu$ are related by $A_f = A_\nu^{(p-1)/2}$, 
if the cooling frequency ($\nu_c$) is above the R-band at 11 min, and 
$A_f=A_\nu^{p/2}$ if $\nu_c$ is below the R-band.
 For $\E_{52}=1$, $\epsilon_e=0.5$, $A_\nu=4$ \& $p=2.2$ we find 
$n_0=61$ cm$^{-3}$, $\Gamma_d=250$, and $\epsilon_B=8\times10^{-6}$.

\subsubsection{Parameters for $s=2$ model from late time Optical data}

The deceleration time for $s=2$ is given by (see eq. \ref{tda})
\begin{equation}
t_d = 0.08 (1+z) {\E_{52}f_\eta \over A_* \Gamma_{d_2}^4} \quad {\rm s},
\label{tda2}
\end{equation}
where $A_* = A/5$x$10^{11}$g cm$^{-1}$ \& $f_\eta=\max(1,\sqrt{2\eta}/3)$. 
Since the peak flux at 11 min is 0.2 $A_f$ mJy, we find using 
equation (\ref{fnua}), the FS peak flux at an earlier time to be
\begin{equation}
f_{\nu_p,f} = 5.4\, A_f t^{-1/2}\quad {\rm mJy},
\label{fnu2}
\end{equation}
and
\begin{equation}
A_*^{3/2}\epsilon_{B_{f_{-4}}}^{1/2} t_d^{1/2}\Gamma_{d_2}^2 = 
   4.5\times10^{-3}A_f
\label{cons21}
\end{equation}
Taking the synchrotron peak frequency at 11 min to be 4.7x10$^{14}
A_\nu^{-1}$ Hz, and substituting this into equation (\ref{numa}) we obtain 
for $s=2$
\begin{equation}
A_*^{1/2} \epsilon_e'{^2}\epsilon_{B_{f_{-4}}}^{1/2}\Gamma_{d_2}^2 t_d^{1/2}=
   5\times10^{-3} A_\nu^{-1},
\label{cons22}
\end{equation}
and the time evolution of $\nu_{if}$ is same as in equation (\ref{num0}).

 From equations (\ref{cons21}) and (\ref{cons22}) we obtain
\begin{equation}
 A_* = 0.9\,\epsilon_e'{^2} (A_\nu A_f),
\label{Astar}
\end{equation}
which when substituted into equation (\ref{tda2}) gives the LF at deceleration
\begin{equation}
\Gamma_{d_2} = 0.7\, \E_{52}^{1/4}t_d^{-1/4}f_\eta^{1/4}\epsilon_e'^{-1/2}
   (A_\nu A_f)^{-1/4}.
\label{gamd2}
\end{equation}
Combining equations (\ref{cons22}), (\ref{Astar}) and (\ref{gamd2}) we
obtain
\begin{equation}
 \epsilon_{B_f}(t=11min) = 1.7\times10^{-8} A_\nu^{-2}f_\eta^{-1}\E_{52}^{-1} 
     \epsilon_e'{^{-4}}.
\label{epsilB2}
\end{equation}
We see from these equations that for $s=2$, $A_\nu$ can be as large
as 10-20, and yet give acceptable values for various parameters.

\section{Reverse Shock}

The emission from reverse shock (RS) in gamma-ray bursts is discussed by 
a number of authors e.g. Panaitescu \& \Meszaros (1998), Sari \& Piran (1999), 
Kobayashi (2000), Piran (2000). A particularly important parameter
that determines the behavior of RS is the thickness of the shell of material 
or ejecta that carries the relativistic energy of the explosion. We have
parametrized the ejecta thickness as $\eta R/\Gamma_0^2$ in lab frame; for
a shell whose thickness is dominated by expansion at deceleration radius
we expect $\eta\sim 1$, otherwise the thickness is determined by the
duration of the central engine, and $\eta$ could be much larger than unity
at $t_d$. We calculate RS emission for a range of $\eta$ between 0.5 and 10.
Fortunately the main conclusions of this work for GRB 021211 remain unchanged
even for a larger range of $\eta$.

 At the deceleration radius $R_d$, the ratio of the thermal energy of 
protons in the reverse shock (RS) region, $\gamma_{p,r}$, to that in the FS, 
$\gamma_{p,f}$, is (see fig. 1)
\begin{equation}
{\gamma_{p,r}\over \gamma_{p,f}} \approx {1\over 4\Gamma_0} \left({n_{ej}\over 
    \Gamma_0^2 n}\right)^{-0.7} \approx {\left[(3-s)\eta\right]^{1.4}\over 
    4\Gamma_0}={\left[(3-s)\eta\right]^{0.9}\over \sqrt{32}\,\Gamma_d}.
\label{gamr}
\end{equation}
The first part in the above equation is valid only for 
$0.01\lta n_{ej}/\Gamma_0^2 n \lta 100$; for $n_{ej}/\Gamma_0^2 n\ll 0.01$, 
it can be shown that $\gamma_{p,r}/\gamma_{p,f} \simeq (n_{ej}/n)^{-0.5}$. 
In deriving the second part of this equation we made use of (\ref{nejRa})
for the density of the ejecta at $R_d$ --- $n_{ej}(R_d)$. 
It should be noted that the thermal energy per proton in RS is 
$\sim m_pc^2\eta^{0.9}/32^{1/2}$, and so protons are not heated to a 
relativistic temperature in the reverse shock.

The pressure continuity across the contact discontinuity surface, 
which separates forward and reverse shocks, implies that the magnetic
field strength in RS and FS are equal, provided that $\epsilon_B$ is
the same behind both shocks. However, one might expect $\eB$ in the 
RS ($\eBr$) to be different from the value in FS ($\eBf$). Then the 
synchrotron peak frequency in the RS is 
\begin{equation}
  \nu_{ir}(t_d) = \nu_{if}(t_d) \left(\frac{\eBr}{\eBf}\right)^{1/2} 
    \left({\gamma_{p,r}\over \gamma_{p,f}}\right)^2_{t_d} =
    \nu_{if}(t_d) { \left[(3-s)\eta\right]^{1.8}
   \over 32\Gamma_d^2}  \left(\frac{\eBr}{\eBf}\right)^{1/2} .
\label{numr1}
\end{equation}
This can be written out explicitly as follows 
\begin{equation}
  \nu_{ir}(t_d) = {q m_p^2 \eBr^{1\o 2} {\epsilon'_e}^2 A^{1\o 2} R_d^{-{s\o2}} 
      \Gamma_d^2[(3-s)\eta]^{1.8} \over (128\pi)^{1/2}(1+z) m_e^3},
\label{numr}
\end{equation}
or
\begin{equation}
\nu_{ir}(t_d) = \epsilon_{B_{r_{-4}}}^{1/2}\epsilon_e'{^3} (A_\nu A_f)^{1/2}
      \eta^{1.8}\times\Biggl\{   {\parbox{5.0cm}{\begin{eqnarray}
        1.1\times10^{16} t_d^{-3/4} \;{\rm Hz} \qqqquad{\rm s=0}\nonumber \\
        5.1\times10^{15} t_d^{-1} \quad{\rm Hz} \qqqquad{\rm s=2} \nonumber 
      \end{eqnarray}  } }
       \Biggl.
\label{numra}
\end{equation}
In deriving this last equation we made use of equations (\ref{gamd20}), 
(\ref{n0}), (\ref{Astar}) \& (\ref{gamd2}) for ISM density and LF at 
deceleration.

Since the FS and RS region are moving at same LF, at deceleration, the RS peak 
synchrotron flux is equal to the FS peak flux times the ratio of
number of electrons in the ejecta to the swept-up electrons in
the surrounding medium up to $R_s$; this ratio is equal to $\Gamma_0/(3-s)
\eta=[2/(3-s)\eta]^{1/2} \Gamma_d$. Thus, the RS peak flux is
\begin{equation}
  f_{\nu_p,r} (t_d)= f_{\nu_p,f}(t_d) \left[{2(\eBr/\eBf)\over(3-s)\eta}\right]^{1/2} 
   \Gamma_d,
\label{fnur1}
\end{equation}
or 
\begin{equation}
  f_{\nu_p,r}(t_d)= {(3\eBr A)^{1\o 2}q^3\E (1+z)^{s\o 2}\over m_e m_p c^3
     d_L'^2 \Gamma_d^{s-1} (4ct_d)^{s\o 2} [\pi(3-s)\eta]^{1\o 2} }.
\label{fnur}
\end{equation}
Using equations (\ref{gamd20}), (\ref{n0}), (\ref{Astar}) \& (\ref{gamd2}) this
equation reduces to
\begin{equation}
f_{\nu_p,r}(t_d) = \epsilon_{B_{r_{-4}}}^{1/2}\epsilon_e'{^{3/2}} 
     (A_\nu A_f)^{3/4}\E_{52}^{3/4}\eta^{-1/2}\times\Biggl\{
         {\parbox{6.6cm}{ \begin{eqnarray}
        2.9\times10^{3} t_d^{-3/8} \quad{\rm mJy}\qqqquad{\rm s=0}\nonumber \\
        3.8\times10^{4} t_d^{-3/4} \quad{\rm mJy}\qqqquad{\rm s=2}\nonumber 
      \end{eqnarray}  } }
       \Biggr.
\label{fnupra}
\end{equation}

The RS synchrotron self-absorption frequency is
\begin{equation}
 \nu_{Ar}^2(t_d) \left({\nu_{ir}\over\nu_{Ar}}\right)^\alpha = 
    {q^3 E\left[ 6\pi\epsilon_B A (1+z)^s \right]^{1/2} \over 8\pi m_e 
    m_p^2 c^3 \epsilon_e' [(3-s)\eta]^{1.4} (4c t_d)^{s+4\o2}\Gamma_d^{s+2}},
\label{nuARS}
\end{equation}
where $\alpha=1/3$ if $\nu_{ir}>\nu_{Ar}$, and $p/2$ otherwise.
Using equations (\ref{gamd20}), (\ref{n0}), (\ref{Astar}) \& (\ref{gamd2}) this
can be rewritten as
\begin{equation}
\nu_{Ar}(t_d) \left( {\nu_{ir}\over\nu_{Ar} }\right)^{\alpha/2} = 
     \epsilon_{B_{r_{-4}}}^{1/4}\epsilon_e' (A_\nu A_f)^{3/4}
        \eta^{-0.7}\times\Biggl\{ 
        {\parbox{6.0cm}{\begin{eqnarray}
        5.5\times10^{13} t_d^{-5/8} \;{\rm Hz}\;\qqqquad{\rm s=0}\nonumber \\
           1.1\times10^{15} t_d^{-1}\quad{\rm Hz}\;\qqqquad{\rm s=2}\nonumber 
      \end{eqnarray}  } }
       \Biggr.
\label{nuARSa}
\end{equation}
 
\subsection{Time dependence of radiation from reverse shock}

The time dependence for $\nu_{ir}$ \& $f_{\nu_p,r}$ is determined by
the evolution of magnetic field and electron thermal energy in the 
reverse shock. Electrons in the ejecta cease to be heated after the 
passage of the RS, and their energy decreases with time as a result of 
adiabatic expansion. If electrons continue to exchange energy with
protons, and the fraction of thermal energy in electrons, $\epsilon_e$,
is time independent, then electron thermal LF decreases as
\begin{equation}
\gamma_e \propto (R^2\delta R)^{-{2-\epsilon_e\o 3}}\propto 
        t^{-{2(2-\epsilon_e)\o 3(4-s)}},
\end{equation}
where $\delta R$ is the comoving shell thickness which is a weak function 
of time for sub- or mildly-relativistic RS, and $R$, the radius of the 
ejecta, increases with time as $t^{1/(4-s)}$; $\epsilon_e=1$ if electrons
and protons are decoupled. 

The magnetic field, frozen in the ejecta, decreases as $B'\propto 
(R\delta R)^{-1}\propto t^{-1/(4-s)}$ if the field is transverse; 
a longitudinal field decreases as $t^{-2/(4-s)}$, therefore any non-zero
transverse field will become the dominant component at large distances.

The synchrotron injection frequency and the peak flux decrease as
\begin{equation}
\nu_{ir} \propto B' \gamma_e^2 \Gamma \propto t^{-{31-3s-8\epsilon_e\over
    6(4-s)}}, \quad {\rm and} \quad f_{\nu_p,r}\propto N_e \, B'\Gamma\propto
    N_e t^{-{5-s\over 2(4-s)}},
\label{nuT}
\end{equation}
where $N_e$ is the total number of ``radiating'' electrons in the ejecta.

The cooling frequency decreases at the same rate as $\nu_{ir}$; the 
decline is faster if radiative losses dominate over adiabatic losses.
The number of electrons radiating in an observer band might have a
non-trivial time independent if the magnetic field is not constant 
across the ejecta -- electrons in a region of higher magnetic field 
will lose energy at a higher rate and their radiation drops below the 
observed band sooner than electrons in lower magnetic field region. 
This together with uncertainty with the evolution of $\gamma_e$ -- which 
depends upon coupling between electrons \& protons -- and the unknown 
energy density \& LF structure of the ejecta, makes it difficult to 
calculate with confidence the power-law decay index for flux from 
reverse shock. 

In order to fit the data for GRB 021211 what we do instead is to work 
backwards from the observed lightcurve slope and determine the decay of 
synchrotron frequency -- which depends on both $B'$ and $\gamma_e$
and so its time dependence is more uncertain than the peak flux which depends
on $B'$ alone -- that we need to calculate the cooling frequency at 11 minutes
after the explosion to make sure that it is above the optical R-band.

Let us consider the time dependence for injection frequency and peak flux 
to be $t^{-\alpha_\nu}$ \& $t^{-\alpha_f}$ respectively. The flux above the
synchrotron peak decays as $t^{-[2\alpha_f+(p-1)\alpha_\nu]/2}$ 
($p$ is electron energy power-law index). The observed decay for GRB 021211
was $t^{-1.8}$, which we use to determine $\alpha_\nu$; we assume that 
$\alpha_f$ is as given in equation (\ref{nuT}), but allow for a small 
deviation when fitting the observed data. With $\alpha_\nu$ and $\alpha_f$
thus determined, we find the time dependence of $B'$ \& $\gamma_e$

\begin{equation}
B' \propto t^{-\alpha_f + {3-s\over 8-2s}}, \qquad \gamma_e\propto
    t^{-{\alpha_\nu - \alpha_f\over 2}},
\end{equation}
that we use to calculate absorption and cooling frequencies and flux 
as a function of time.
 
\subsection{Compton Parameter \& Cooling Frequency}

The comoving frame timescale for an electron of energy $m_e c^2\gamma_e$ to 
cool as a result of synchrotron and inverse Compton emission is 
\begin{equation}
t'_c = {6\pi m_e c\over \sigma_T B'^2 \gamma_e (1+Y)} = 
  {3m_e R_d^s\over 16\sigma_T c \epsilon_B \gamma_e \Gamma_d^2 A (1+Y)},
\end{equation}
where $Y$ is the Compton $Y$ parameter, prime denotes comoving
quantity, and the cooling is considered at the deceleration time. 
At deceleration, when $t'_c = 4 t_d \Gamma_d/(1+z)$, the  
electron cooling LF, $\gamma_c$, defined by the equality of the
radiative and dynamical timescales, is 
\begin{equation}
\gamma_c(t_d) = {3\pi m_e c (1+z)\over 2\sigma_T B'^2 t_d \Gamma_d (1+Y)}=
  {3 m_e \Gamma_d^{2s-3}\over 16\sigma_T\epsilon_B A (1+z)^{s-1}
  (4ct_d)^{1-s} (1+Y)}
\end{equation}
Substituting for $A$ and $\Gamma_d$ from equations (\ref{gamd20}), (\ref{n0}), 
(\ref{Astar}) \& (\ref{gamd2}) we obtain
\begin{equation}
\gamma_c(t_d) = {\E_{52}^{1/4}\over \epsilon_{B_{-4}}{\epsilon_e'}^{5\o 2}
       (A_\nu A_f)^{5\o 4} (1+Y)} \times\Biggl\{
        {\parbox{4.0cm}{\begin{eqnarray}
        513\, t_d^{1/8} \qqqquad{\rm s=0}\nonumber \\
           22\, t_d^{3/4} \qqqquad{\rm s=2}\nonumber 
      \end{eqnarray}  } }
       \Biggr.
\label{gamca}
\end{equation}

The cooling frequency $\nu_c$, defined as the synchrotron frequency for 
electrons with LF $\gamma_c$, is
\begin{equation}
\nu_c(t_d) = {qB' \gamma_c^2\Gamma_d\over 2\pi m_e c (1+z)} = 6.1\times 10^{-5}
     {c^2 (4c t_d\Gamma_d^2)^{3s-4\o 2}\over(A\,\epsilon_{B_{-4}})^{3/2}(1+Y)^2
      (1+z)^{(3s-2)/2}  } \;\,{\rm Hz},
\label{nuc}
\end{equation}
which can be rewritten by substituting for $A$ and $\Gamma_d$
\begin{equation}
\nu_c(t_d) = {\E_{52}^{1/2}\over \epsilon_{B_{-4}}^{3/2}\epsilon_e'^{4} 
       (A_\nu A_f)^{2} (1+Y)^2} \times\Biggl\{           
        {\parbox{6.6cm}{\begin{eqnarray} 2\times10^{15} (1+z) 
        t_d^{-1/2} \;\quad{\rm Hz}\;\qquad{\rm s=0}\nonumber\\
     9\times10^{13}(1+z)^{-2}t_d^{1/2} \quad{\rm Hz}\qquad{\rm s=2}\nonumber
                     \end{eqnarray}  } }
       \Biggr.
\label{nuca} 
\end{equation}

The cooling frequencies in the reverse and forward shock regions are 
calculated from this equation by substituting appropriate values for 
$\eB$ and $Y$ corresponding to each region. The Compton $Y$ parameter 
is calculated below.

The electron column density in the ejecta, at the deceleration radius,
assuming that the ejecta consists only of protons and electrons, \ie
there are no pairs, is
\begin{equation}
N_{e,r} = {\E\over 4\pi R_d^2 \Gamma_0 m_p c^2} = {(1+z)^2 \E \over 
         64\pi \sqrt{2(3-s)\eta} \,m_p c^4 t_d^2 \Gamma_d^5 }.
\end{equation}
The optical depth of the ejecta to Thomson scattering is
\begin{equation}
\tau_{r} = \sigma_T N_{e,r} = 1.7\times10^{-3} {(1+z)^2 \E_{52} \over 
              \sqrt{(3-s)\eta}\, t_d^2 \Gamma_{d_2}^5 }.
\label{tau}
\end{equation}

The Compton parameter $Y=\tau \overline{\gamma_e^2}$, where
$\overline{\gamma_e^2}$, the mean squared electron LF, for 
$\nu_A < \nu_i < \nu_c$ \& $2<p<3$ is 
\begin{equation}
\overline{\gamma_e^2} = {(p-1)\over (p-2)(3-p)} \gamma_i^2 
   \left({\gamma_c\over \gamma_i}\right)^{3-p}, 
\label{gammae2}
\end{equation}
and $m_e c^2\gamma_i$ is the minimum thermal energy of shock heated electrons.
The Compton parameter in the particular case of $\nu_A < \nu_c < \nu_i$ is
given by
\begin{equation}
Y \simeq \left( {\epsilon_e\over\epsilon_B}\right)^{1/2}.
\label{ycool}
\end{equation}

Substituting (\ref{gamca}) into (\ref{gammae2}) and making use of equations 
(\ref{gamr}) \& (\ref{tau}) we find the Compton $Y$ in the reverse shock 
region
\begin{eqnarray}
Y(1+Y)^{3-p} = {(p-1)m_p^{p-1}\E_{52}^{2-p\o 4}\epsilon_{B_{r_{-4}}}^{p-3}
       {\epsilon_e'}^{7p-12\o 2} (A_\nu A_f)^{5(p-2)\o 4} \eta^{0.9p-1.4}
       \over (p-2)(3-p) m_e^{p-1}32^{p-1\o 2} } \nonumber\\
         \quad\times\Biggl\{
        {\parbox{5.3cm}{\begin{eqnarray}
        2.8\times10^{4}\, 191^{-p} t_d^{-{p-2\o 8}}\qquad{\rm s=0}\nonumber\\
        6.4\times10^{2}\, 22^{-p} t_d^{-3(p-2)\o4}\qquad{\rm s=2}\nonumber 
                     \end{eqnarray}  } } 
       \Biggr.
\label{yrsa}
\end{eqnarray}

The Compton $Y$ substituted back into equation (\ref{nuca}) yields the cooling
frequency in RS. A similar calculation gives $\nu_c$ in the forward shock.

When $\nu_A > \min \{\nu_c, \nu_i \}$, the synchrotron photon flux that is 
scattered by an electron is diminished by self-absorption and $\nu_A$ and 
$\nu_c$ have to be determined by solving a set of coupled equations, as 
described in Panaitescu \& \Meszaros (2000). Some of the cases considered 
for GRB 021211r fall in this more complicated regime, and all of the 
numerical results presented in this paper are obtained by determining 
$\nu_A$ and $\nu_c$ numerically, in a self-consistent manner.

\section{A unified modeling for $\gamma$-ray and afterglow data}
\label{s0}

We apply the results of the last two sections to a systematic analysis of
$\gamma$-ray, optical and radio observations for GRB 021211 and determine
models that are consistent with all data. We discuss the cases of a 
uniform density ISM ($s=0$), and a medium carved out by the progenitor's 
wind ($s=2$) in two separate subsection.

\subsection{A Uniform Density Circumburst Medium (s=0)}

In the next subsection we discuss the early optical \& radio emissions from 
the RS.  We take up the question of what could have produced the 
$\gamma$-ray emission in \S6.1.2 \& radio upper limit in \S6.1.3.

\subsubsection{Optical and radio emissions from reverse shock}

The RS synchrotron injection frequency is a factor of 
$(\gamma_{p,f}/\gamma_{p,r})^2$ smaller compared with the peak 
frequency of the FS emission as long as $\epsilon_B$ is the same in
reverse and forward shocks. The synchrotron injection frequency in the RS,
otherwise, is expected to be about 2x10$^{-2}\eta^{1.8}A_\nu^{-1}
(\eBr/\eBf)^{1/2}$ ev (see eqs. \ref{numra} \& \ref{cons2}). This suggests
that the RS flux in the optical band decreases with time for $t>t_d$.
The extrapolation of the observed flux of 7.2 mJy at 90s, with a
powerlaw decline of $t^{-1.8}$, gives a R-band flux at 5s of 1.3 Jy
or 8.5 mag. So, contrary to claims, GRB 021211 was as bright as 990123 
close to the deceleration time.

The injection frequency in RS for $s=0$ declines with observer time 
approximately as $t^{-1}$, and the cooling frequency too declines as 
$t^{-1}$ or faster. After the passage of the reverse shock, which takes 
place on the deceleration time
scale of a few seconds for GRB021211, electrons are no longer accelerated and
there is no emission from RS at a frequency greater than the cooling
frequency ($\nu_{cr}$).
The R-band flux from GRB 021211 is observed to decline as $t^{-1.8}$ for
11 minutes and then the decline slows down to $t^{-0.8}$. This suggests
that the RS emission lasts for at least 11 minutes in the R-band, and
therefore $\nu_{cr}$ at deceleration should be $\gta 10^{17}$ Hz. 

The inverse Compton parameter and the cooling frequency are determined from
equations (\ref{yrsa}) and (\ref{nuca}). For $p=2.5$ these quantities at
the deceleration time are
\begin{equation}
Y(1+Y)^{0.5} = 2\times10^3\epsilon_e'^{2.75}\epsilon_{B_{r_{-4}}}^{-{1\o2}}
      \eta^{0.85}A_\nu^{35\o32}\E_{52}^{-{1\o8}} t_d^{-{1\o16}},
\label{yrss0}
\end{equation}
and
\begin{equation}
\nu_{cr} = 1.5\times10^{11} {\E_{52}^{1/2} \over\epsilon_e'^{7.7}
      \epsilon_{B_{r_{-4}}}^{0.83}\eta^{1.1}A_\nu^{5}t_d^{2/5} }\quad
     {\rm Hz},   
\label{nucs0}
\end{equation}
so long as $Y\gta1$. For $Y\ll1$ the cooling frequency is obtained by
setting $Y=0$ in equation (\ref{nuca}).

The requirement that $\nu_{cr}(t_d)>10^{17}$Hz -- in order to have 
non-zero flux in the R-band from RS for $\sim 11$min -- provides an 
upper limit on $\epsilon_e'$ given below
\begin{equation}
\epsilon_e'\lta {0.175\,\,\E_{52}^{1/15.4}\over \epsilon_{B_{r_{-4}}}^{1/9}
     A_\nu^{2/3}\eta^{1/7} t_d^{1/20}}.
\end{equation}
Substituting this into (\ref{numra}) \& (\ref{fnupra}) we find the injection 
frequency and peak flux from the RS
\begin{equation}
\nu_{ir}(t_d)\lta 6\times10^{13} \epsilon_{B_{r_{-4}}}^{1/6}\E_{52}^{1/5.1}
    A_\nu^{-1.1}\eta^{1.37}t_d^{-0.9}\quad{\rm Hz},
\end{equation}
and
\begin{equation}
f_{\nu_p,r}(t_d)\lta 219\, \epsilon_{B_{r_{-4}}}^{1/3}\E_{52}^{0.85}
    A_\nu^{7/20} \eta^{-0.7}t_d^{-0.45}\quad{\rm mJy}.
\end{equation} 
The flux in the R-band ($\nu_R=4.95\times10^{14}$Hz) at deceleration time is
\begin{equation}
f_{\nu_R}(t_d) \lta 46\, \epsilon_{B_{r_{-4}}}^{0.46}\E_{52}
    A_\nu^{-7/16} \eta^{0.3}t_d^{-1.1}\quad{\rm mJy},
\end{equation}
whereas the flux in R-band at 90s is $f_{\nu_R}(t_d)(t/t_d)^{-1.8}$, if 
$\nu_{ir}(t_d)<\nu_R$, otherwise the flux is given by 
\begin{equation}
f_{\nu_R}(t=90s) = f_{\nu_p,r}(t_d) (90s/t_R)^{-1.8} (t_R/t_d)^{-\alpha_f},
\end{equation}
where $t_R$ is the larger of $t_d$ \& the time when 
$\nu_{ir}(t)=4.95\times10^{14}$ Hz (R-band frequency), and $\alpha_f\sim1$
is the power-law decay index for the peak-flux i.e. $f_{\nu_p,r}(t)\propto 
t^{-\alpha_f}$. The above equations for R-band flux are applicable when 
the synchrotron-self-absorption frequency is less than $\nu_R$ which is 
indeed the case as (see eq. \ref{nuARSa}).

Since $\nu_{ir}(t_d)<4.95\times10^{14}$Hz, the R-band flux at 90s is found
to be $0.014\,\epsilon_{B_{r_{-4}}}^{0.46}\E_{52}A_\nu^{-7/16}\eta^{0.3}
t_d^{0.7}$ mJy. A R-band flux of 7.2 mJy, as observed for 021211, requires
$\epsilon_{B_{r}}\sim 10^{-2}$, $\E_{52}\sim 20$ \& $A_\nu\sim\eta$ 
($t_d\sim 3 s$). Figure 2 shows the allowed parameter space which 
satisfies the R-band flux at $t\ge90$s. Note that $\epsilon_{B_r}/
\epsilon_{B_f}\gta 10^3$, for the allowed parameter space. This result
is consistent with the analytical calculation presented above.

 Fox et al. (2003) reported that the flux at 8.5GHz, 0.1 day after the burst,
was less than 110$\mu$Jy. This frequency is above the self-absorption 
frequency, and the flux from RS is a few times larger than this upper 
limit for the parameter space in fig. 2, for $t_d\sim3$s. Diffractive 
interstellar scintillation can decrease the flux at 8.5GHz at this 
early time by a factor of a few thereby providing consistency with 
the reported flux upper limit. We note that the radio flux would exceed
the observational limit by almost an order of magnitude if the time for
RS crossing were taken to be 30s, almost independent of the details of
RS model, which suggests that the shock crossing time is approximately
equal to the burst duration of 2--4 s.

\subsubsection{$\gamma$-ray emission during the GRB}

 The injection frequency in FS at the deceleration time ($t_{d}=2$ s) is 
13.2 $A_\nu^{-1}$ kev (see eq. \ref{num0}), and the peak flux is 
$0.2 A_f$ mJy. To explain the early optical afterglow requires $A_\nu$ to be of
order a few or larger (see fig. 2 \& the discussed in the last subsection),
and therefore the injection frequency and the peak flux in FS at
$t_d$ are $\sim 5$ kev \& 0.4 mJy respectively. The observed values 
for the peak $\gamma$-ray flux during the GRB is about 4 mJy, and the 
peak of $\nu f_\nu$ is at $\sim50$ kev (Crew \etal 2003). Thus, the 
observed peak flux during the GRB is about an order of magnitude larger 
than predicted by the extrapolation of the optical data at 11 min. 
And the observed peak frequency, depending on the value of the cooling
frequency in forward shock, is also about an order of magnitude larger
than the synchrotron peak frequency.

We consider whether synchrotron-self-Compton process in the
reverse or the forward shock might explain the gamma-ray emission.
The peak of $\nu f_\nu$ for inverse Compton scattered synchrotron photons
occurs at a frequency of 
$\max\{\gamma_i^2, \gamma_c^2\}\times\max\{\nu_i, \nu_c\}$
(see Panaitescu \& Meszaros, 2000). For the reverse shock of GRB021211, 
$\nu_c\sim 1$kev, and $\gamma_{c}\sim 3\times10^2$ (see \S6.1.1), and 
therefore, the IC peak frequency is at $\sim 10^2$Mev, or three order of 
magnitude above the observed peak; the IC flux at 50 kev is about 0.1 mJy. 
For the forward shock, $\nu_i\sim 5$kev \& $\gamma_i\sim 10^4$, and thus
the IC spectrum peaks at a energy $\gta 0.5$Tev; the flux at this energy 
is smaller than the upper limit provided by Milagro (McEnery et al. 2002).

Having eliminated synchrotron-self-Compton process as an explanation for 
the $\gamma$-ray emission from GRB 02121, we turn to synchrotron emission 
from the reverse or the forward shock as a possible mechanism to account
for the observations\footnote{We feel that a simple single peaked
lightcurve for 021211 should not require internal shocks to produce 
$\gamma$-ray emission, which were invoked to explain multi-peaked and 
highly fluctuating GRBs.}.

The synchrotron emission from RS can have $\nu_c\sim 50$kev provided that
we consider a small value for $\epsilon_e'\sim 0.04$. The flux at 50 kev
can be calculated directly from the observed optical R-band flux and is
estimated to be about 4 mJy -- consistent with the observed $\gamma$-ray
flux. This would have been a very economical and elegant explanation for 
all the observations for 021211 from $\gamma$-ray to radio frequencies. 
However, this possibility is, unfortunately, ruled out by the observed 
spectral slope of 0.4 ($f_\nu\propto \nu^{0.4}$) below the 50 kev peak
(Crew et al. 2003), whereas the RS synchrotron model predicts a 
spectral power-law index of $-(p-1)/2\sim -0.5$.

As we discussed earlier, synchrotron emission from the forward shock cannot 
account for the gamma-ray observations as long as we take $\epsilon_B$ in 
FS at deceleration to be same as it is at 11 minutes. Having ruled out
all possibilities for producing $\gamma$-rays in a standard external shock
(for $s=0$), we now relax the assumption of time independent $\epsilon_B$ 
in the FS. Our goal is to find a solution where $\nu_c\sim\nu_i\sim 50$kev, 
and the peak flux is 4 mJy; note that $\nu_c\sim\nu_i$ is required by the 
observed low energy spectral index\footnote{For any other ordering of $\nu_c$ 
\& $\nu_i$ the spectral index below the peak of $\nu f_\nu$ will be close 
to -0.5.}.

For a time independent $\epsilon_B$ in the forward shock $\nu_{if}(t_d)\approx
37.3\, t_d^{-3/2} A_\nu^{-1}$ Kev (see eq. \ref{num0}), and the peak flux is 
$0.2\, A_f$mJy. Therefore, the flux at 50 kev is $\sim 0.2\,A_f 
(\nu_{if}/50)^{(p-1)/2} \sim 0.16\,t_d^{-9/8}$mJy; note that 
$A_f=A_\nu^{(p-1)/2}$. This flux is too small by a factor of about $10^2$ 
to satisfy the $\gamma$-ray observation for 021211. The only way out of 
this difficulty is to consider $\epsilon_{B_f}$ to be larger at the deceleration
time by a factor of $\sim 100$ than its value at 11 minutes. Since the flux 
at $\nu_i<\nu<\nu_c$ scales as $\epsilon_B^{(p+1)/4}$, and the synchrotron 
injection frequency $\nu_i\propto\epsilon_B^{1/2}$, a larger $\epsilon_B$ 
by a factor $10^2$ will increase the flux at 50 kev by two orders of magnitude 
\& increase $\nu_{if}$ by a factor 10, thereby simultaneously satisfying
the observed $\gamma$-ray flux and the peak frequency requirements.
 The rest of this section is devoting to the calculation of cooling 
frequency in FS at the deceleration time to make sure that 
a larger value of $\epsilon_B$ at $t_d$ is compatible with the 
requirement of $\nu_{cf}\sim50$kev.

The optical depth to Thomson scattering in forward shock at the 
deceleration time is
\begin{equation}
\tau_{f} = {\sigma_T n_0 R_d\over 3 m_p} = 8.9\times10^{-6} {\epsilon_e'{^3}
      (A_f A_\nu)^{3/2} t_d^{1/4}\over (1+z)\E_{52}^{1/2} },
\label{tauf}
\end{equation}
where $R_d=4ct_d\Gamma_d^2/(1+z)$ is the deceleration radius, 
and the last equality was obtained by making use of equations 
(\ref{gamd20}) \& (\ref{n0}) to eliminate $\Gamma_d$ (LF at deceleration) \&
$n_0$ (ISM density). The Compton parameter $Y=\tau_{f}\overline{\gamma_e^2}$
with $\overline{\gamma_e^2}$ given by equation (\ref{gammae2}) when 
$\gamma_c>\gamma_i$; $\gamma_i=(m_p/m_e)\epsilon_e'\Gamma_d$ in
the forward shock at $t_d$. Combining these equations we find 
\begin{equation}
{Y\over\gamma_c^{3-p}} = {(p-1)\gamma_i^{p-1}\over (3-p)(p-2)} \tau_{f} =
    4.5\times10^{-6} {(p-1)\Gamma_d^{p-1}\over (3-p)(p-2)} 
    \left({m_p\over m_e}\right)^{p-1} { \epsilon_e'{^{p+2}} (A_f A_\nu)^{3/2} 
     t_d^{1/4}\over \E_{52}^{1/2}}.
\end{equation}
Substituting for $\gamma_c$ from equation (\ref{gamca}) we obtain
\begin{equation}
Y(1+Y)^{3-p} = 4.5\times10^{-6} {(p-1)\over (3-p)(p-2)} 
    \left({m_p\over m_e}\right)^{p-1} {{\epsilon_e'}^{7p-11\o 2}
    (A_f A_\nu)^{5p-9\o 4} \Gamma_d^{p-1} t_d^{1\o 4}\over 
    \epsilon_{B_{f_{-4}}}^{3-p}\left(513\, t_d^{1/8}\right)^{p-3} 
     \E_{52}^{p-1\o 4} }.
\end{equation}
 For $p=2.5$ and $Y\gg1$ the Compton parameter is 
\begin{equation}
Y = 2\times10^{3} \epsilon_{B_{f_{-4}}}^{-1/3} {\epsilon_e'}^{5\o 3}
    (A_f A_\nu)^{1\o 3} t_d^{-1/6}.
\end{equation}
Substituting this into equation (\ref{nuca}) we find the cooling frequency in
forward shock for $z=1$:

\begin{equation}
\nu_{cf} = 7\times10^{9} {\E_{52}^{1\o 2}\over\epsilon_{B_{f_{-4}}}^{5\o 6}
           {\epsilon_e'}^{22\o 3} A_\nu^{14\o 3} t_d^{1/6} } \quad{\rm Hz}.
\label{nucf}
\end{equation}
When $\nu_{cf}<\nu_i$ \& $Y\gg1$, $Y\approx(\epsilon_e/\epsilon_{B_f})^{1/2}$, 
and
\begin{equation}
\nu_{cf} = 2\times10^{11} {(1+z) \E_{52}^{1/2}\over \epsilon_{B_{f_{-4}}}^{1/2} 
           \epsilon_e \epsilon_e'{^4} A_\nu^{7/2} t_d^{1/2} } \quad{\rm Hz}.
\label{nucf1}
\end{equation}
 For $Y\ll1$ the cooling frequency in FS can be obtained by setting $Y=0$ 
in equation (\ref{nuca}).

The peak of the spectrum for 021211 is at $\nu_{cf}\sim\nu_{if}=50\,{\rm kev}
=1.2\times 10^{19}$Hz. Substituting this into equation (\ref{nucf}) gives
\begin{equation}
\epsilon_e' A_\nu^{7\over11}\epsilon_{B_{f_{-4}}}^{5\over44}(t_d) = 
  0.055\,\E_{52}^{3\over44} t_d^{-1\over44}.
\end{equation}
Combining this with equation (\ref{cons2}) -- under the assumption that 
$\epsilon_e'$ is time independent -- and making use of the requirement that 
$\epsilon_{B_f}(t_d)/\epsilon_{B_f}(t=11min)\sim 10^2$
discussed earlier, we find
\begin{equation}
 \epsilon_{B_f}^{6\over11}(t_d) \E_{52}^{14\over11} = 4.3\,A_\nu^{6\over11} 
  t_d^{1\over11}.
\end{equation}
This relation can be satisfied if we consider, for instance, $\E_{52}=20$,
$\epsilon_{B_f}(t_d)\sim 10^{-2}$ \& $A_\nu\sim 1$, and provides a self
consistent solution that accounts for $\gamma$-ray and optical
radiations for GRB 021211.

 Figure 2 shows the parameter space allowed by the $\gamma$-ray flux 
 for 021211 originating in the forward shock. Note that there is a 
range of parameters for which the early and late optical, and 
$\gamma$-ray observations can be simultaneously explained. The general 
requirement, however, is
a large magnetic field in the ejecta, somewhat smaller field
in the forward shock at deceleration, and a much smaller $\epsilon_{B_f}$
at 11 min when the forward shock emission starts to become a dominant 
contributor to the optical flux.

\subsubsection{Radio flux limit at 10 days}

The flux for GRB 021211 at 8.5 GHz, 10 days after the burst, is reported
to be less than 
35 $\mu$Jy (Fox \etal 2003).  The synchrotron peak frequency, which decays
as $t^{-3/2}$, independent of ISM density stratification, is 9.7$A_\nu^{-1}$ 
GHz at 10 days. If the radio band frequency were above the synchrotron 
self-absorption frequency, then the flux at 8.5 GHz should be 0.19 mJy
independent of $A_\nu$, which is a factor of 5.4 larger than the 
observed upper limit.

We consider the possibility that the self-absorption frequency is larger 
than 8.5 GHz by a factor of about 2 thereby reducing the flux below the
observational upper limit. The FS self-absorption frequency, calculated 
using equations (\ref{nuAFS}),  (\ref{cons2})--(\ref{n0}), is
\begin{equation}
\nu_A (\nu_i/\nu_A)^{1/6} = {2.6\times 10^{8}\over (1+z)^{3/2}\epsilon_e'^{1/2}
   }\epsilon_{B_{f_{-4}}}^{1/4}\Gamma_{d_2}^2 n_0^{3/4} t_d^{3/4} t^{-1/4}\;
    {\rm Hz} = 2.4\times10^{10} A_\nu^{p/2} \E_{52}^{-1/2}\epsilon_e'^{1/2}
    t^{-1/4} \; {\rm Hz}.
\end{equation}
In deriving the second equality we have set $z=1$. Thus, $\nu_A\sim 
1.2\times10^8$ Hz at 10 days for $A_\nu=4$, $\E_{52}=10$ \& $\epsilon_e'=0.03$. 
We see that $A_\nu$ should be $\sim 100$ in order that $\nu_A\sim20$ GHz,
 and the flux in 8.5 GHz band at 10 days is below 35 $\mu$Jy. However,
for $A_\nu\sim 100$, $n_0\sim 2\times10^6$ cm$^{-3}$ \&
$\epsilon_B\sim10^{-8}$ (see eqs. \ref{gamd20} \& \ref{n0}); these values 
are in contradiction with the 
requirement that $n_0\sim10^{-2}$ cm$^{-3}$ \& $\epsilon_{B_f}(t=11min)
\sim10^{-4}$ in order to produce $\gamma$-ray emission in external shock 
(see fig. 2).

A decrease in the density with $r$, such as for $s=2$ medium, can reconcile 
the radio flux limit at 10 days. However, it is very difficult to
produce $\gamma$-ray radiation, as seen in 021211, for $s=2$ density 
stratification (see \S6.2.2).

A loss of explosion energy by a factor of about 10 between 11 min and 10 
days would also reduce the radio flux to a value below the observational 
upper limit (the peak flux \& synchrotron peak frequency are proportional 
to $\E$ \& $\E^{1/2}$ respectively). This requirement is, however, 
inconsistent with small $\epsilon_B$ at late times ($t>11$min), and 
the fact that $\nu_c>\nu_i$ at all times.

 Yet another way that the radio flux at 10 days can be reduced
is by requiring the jet break time to be less than 10 days.
The isotropic equivalent of energy in 021211, estimated from early
optical data (\S6.1.1), is 2x10$^{53}$ erg, which suggests 
 jet opening angle to be about 8$^o$ (energy in GRBs is
found to be narrowly clustered around 10$^{51}$ erg -- Panaitescu \& Kumar 
2002, Frail et al. 2002, Piran et al. 2002). For $n_0\sim10^{-2}$ 
cm$^{-3}$ (see fig. 2), we expect the jet break time to be about 
10 days -- which is consistent with the lack of a clear break in optical
light-curve -- and this can reduce the radio flux by a factor of a few. 
Another factor of 2 decrease could come from a decrease in 
$\epsilon_B$ by a factor of 2 between 11 minutes \& 10 days\footnote{
The decline of optical lightcurve as $t^{-0.82}$ together with the
optical spectrum of $\nu^{-0.9}$ limits the decline of $\epsilon_B$
between 11min and 10 days to be less than a factor of $\sim3$.}.
These effects together could then reconcile the late time radio flux in the
model considered here with the observational upper limit. 

\subsubsection{Milagro limit}

Milagro reported an upper limit of $4\times 10^{-6}$ erg cm$^{-2}$ on 
the 0.2-20 Tev fluence for GRB 021211 (McEnery \etal 2002). We discuss
whether the solutions we have found for early optical and $\gamma$-ray
radiations, shown in fig. 2, respect this limit.

The peak of the inverse-Compton in reverse shock is at $\nu_c\gamma_c^2
\sim 10^2$Mev, and the fluence in Milagro band is estimated to 
be $\sim 10^{-8}$erg cm$^{-2}$.  The IC emission in the forward shock 
peaks at $\nu_i\gamma_i^2 \sim 10$Tev, and the fluence in 0.2-20 Tev 
is $\sim Y \nu_i f_{\nu_i} t_d \sim 5\times10^{-7}$erg cm$^{-2}$ -- 
smaller than the Milagro upper limit by an order of magnitude.

\subsubsection{A summary of results for a uniform-density medium}

A uniform density medium cannot simultaneously explain the
R-band afterglow emission after 11 minutes (FS emission), 
and before 11 minutes (presumably from the RS) unless
the energy density in magnetic field in the RS is at least 
a few hundred times larger than the magnetic energy density in 
FS. Moreover, for $s=0$ and an external shock origin for GRB 021211,
the $\gamma$-ray emission can be produced via the synchrotron process in
the forward shock provided that the magnetic field parameter
($\epsilon_B$) in FS at deceleration is larger by a factor 
$\sim10^2$ compared with the value at 11 minutes, i.e., it requires
a time dependent $\epsilon_{B_f}$ during the first few minutes. To satisfy
the upper limit on radio flux at 10 days seems to require a combination of jet
break at about 10 days and a decline of $\epsilon_B$ by a factor of
$\sim2$ between 11 min and 10 days.

\subsection{Pre-Ejected Wind Circumburst Medium (s=2)}
\label{s2}

We consider early optical emission from reverse shock in the next
sub-section, $\gamma$-ray emission in \S6.2.2, and radio data in \S6.2.3.

\subsubsection{Optical emission from reverse shock for s=2}

The synchrotron injection frequency and the flux at the peak of the 
RS spectrum are given by equations (\ref{numra}) \& (\ref{fnupra}).
For $p=2.5$, $\E_{52}=10$, $A_\nu=4$, $A_f=A_\nu^{(p-1)/2}=2.8$, 
$t_d=2$s, $\eta=4$, and $\epsilon_e'=0.05$,
we find $\nu_{ir}=1.3\times10^{13}\epsilon_{B_{r_{-4}}}^{1/2}$Hz, and 
$f_{\nu_p,r}=7.4\epsilon_{B_{r_{-4}}}^{1/2}$Jy; note that 
$\Gamma_d=234$, $A_*=0.03$ \& $\epsilon_{B_f}(t=11min)=1.7\times10^{-5}$
for this choice of parameters. The resulting flux in the R-band at 90s is 
$\lta 0.5$mJy or a factor 15 smaller than the observed value if we
take $\epsilon_{B_{r_{-4}}}=1$. For $\epsilon_{B_{r_{-4}}}\sim 10^2$
the flux agrees with the optical observation, provided that absorption 
frequency is below the R-band and the cooling frequency is sufficiently high,
$\nu_{cr}\sim 10^{17}$Hz, so that electrons in the RS continue to
radiate in the R-band for $\sim11$ minutes. We look into these requirements
below, and determine the set of parameters that satisfies optical observations.

The RS synchrotron self-absorption frequency at deceleration time is given 
in equation (\ref{nuARSa}). For the parameters considered above the 
self-absorption frequency is $\sim10^{14}$ Hz, below the R-band, and 
does not affect the optical flux.

The calculation of cooling frequency proceeds in the same manner as
for the $s=0$ case considered in the previous section. The inverse Compton 
parameter and the cooling frequency are determined from
(\ref{yrsa}) \& (\ref{nuca}). For $p=2.5$ \& $s=2$ these 
quantities at the deceleration time are

\begin{equation}
Y(1+Y)^{0.5} = 1.1\times10^4\epsilon_e'^{2.75}\epsilon_{B_{r_{-4}}}^{-1/2}
      \eta^{0.85}A_\nu^{35/32}\E_{52}^{-1/8} t_d^{-3/8},
\label{yrss2}
\end{equation}
and
\begin{equation}
\nu_{cr} = 9.5\times10^{7} {\E_{52}^{1/2} t_d \over\epsilon_e'^{7.7}
      \epsilon_{B_{r_{-4}}}^{0.83}\eta^{1.1}A_\nu^{5} }\quad
     {\rm Hz},   
\label{nucs2}
\end{equation}
so long as $Y\gta1$. For $Y\ll1$ the cooling frequency is given by
setting $Y=0$ in equation (\ref{nuca}).

The requirement that $\nu_{cr}(t_d)>10^{17}$Hz -- in order to have 
non-zero flux in the R-band from RS for $\sim 11$min -- provides an 
upper limit on $\epsilon_e'$ 

\begin{equation}
\epsilon_e'\lta {0.067\,\E_{52}^{1/15.4} t_d^{1/7.7}\over
     \epsilon_{B_{r_{-4}}}^{1/9} A_\nu^{2/3}\eta^{1/7} }.
\end{equation}
Substituting this into (\ref{numra}) \& (\ref{fnupra}) we find the injection 
frequency and peak flux in the RS
\begin{equation}
\nu_{ir}(t_d)\lta 1.5\times10^{12} \epsilon_{B_{r_{-4}}}^{1/6}\E_{52}^{1/5.1}
    A_\nu^{-1.1} \eta^{1.37}t_d^{-0.6}\quad{\rm Hz},
\end{equation}
\begin{equation}
 f_{\nu_p,r}(t_d)\lta 660\,\epsilon_{B_{r_{-4}}}^{1/3}\E_{52}^{0.85}
    A_\nu^{7/20} \eta^{-0.7}t_d^{-0.55}\quad{\rm mJy},
\end{equation} 
and the flux in R-band at deceleration time is
\begin{equation}
 f_{\nu_R}(t_d) \lta 8.5\,\epsilon_{B_{r_{-4}}}^{0.46}\E_{52}
    A_\nu^{-7/16} \eta^{0.3}t_d^{-1.1}\quad{\rm mJy}. 
\end{equation}
The flux in R-band at 90s is therefore about $2.6\times10^{-3}\,
\epsilon_{B_{r_{-4}}}^{0.46}\E_{52}A_\nu^{-7/16}\eta^{0.3}t_d^{0.7}$ mJy. 
This is a factor 2 smaller than the observed value of 7.2 mJy, even when
we take $\epsilon_{B_{r_{-4}}}=10^4$, $\E_{52}=20$, and $A_\nu=1$. So,
to obtain the desired optical flux at 90s, in a pre-ejected 
wind circum-burst-medium model, requires very extreme, and perhaps 
unphysical, parameters. More accurate numerical 
calculations support this conclusion.

\subsubsection{$\gamma$-ray emission for s=2 medium}

The arguments against synchrotron self-Compton in reverse or forward
shock as an explanation for the $\gamma$-ray observations for GRB 021211,
we suggested for $s=0$, also apply to $s=2$. So we once again turn to
synchrotron radiation in forward shock, the most likely mechanism for
021211, to explain the $\gamma$-ray observations. 

For a time independent $\epsilon_B$ in the forward shock 
the synchrotron {\bf injection} frequency at the deceleration
is $37.3\, A_\nu^{-1}t_d^{-3/2}$kev -- a factor $\sim 4$ smaller than
the observed $\nu_{peak}$ --  and the peak flux is $5 A_f 
t_d^{-1/2}$mJy. The flux at 50 kev is $\sim 5 A_f t_d^{-1/2}
 (\nu_{if}/50)^{(p-1)/2} \sim 4 t_d^{-13/8}$mJy, which is also a factor
$\sim4$ smaller compared with the observed flux. Removing these discrepancies
requires $\epsilon_{B_f}$ at $t_d$ to be larger by an order magnitude
than the value at 11 minutes.

As discussed in last section we require $\nu_{cf}\sim\nu_{if}\sim50$ kev
in order to be consistent with 021211 $\gamma$-radiation. We compute
the cooling frequency below to determine whether this condition can 
be satisfied for $s=2$.

The equation for Compton parameter in FS for $s=2$, when $\gamma_c>\gamma_i$, 
is derived in the same way as the $s=0$ case considered in \S6.1.2, and 
is found to be 
\begin{equation}
Y(1+Y)^{3-p} = 7.4\times10^{-4} {(p-1)\over (3-p)(p-2)} 
    \left({m_p\over m_e}\right)^{p-1} {\epsilon_e'{^{7p-11\o2}}
    (A_f A_\nu)^{5p-9\o4} \Gamma_d^{p-1}\over 
    \epsilon_{B_{f_{-4}}}^{3-p}\left(22\,t_d^{3/4}\right)^{p-3} 
     \E_{52}^{p-1\o4} t_d^{1/2} }.
\end{equation}
 For $p=2.5$ and $Y\gg1$ the Compton parameter is 
\begin{equation}
Y = 8.7\times10^{3}\epsilon_{B_{f_{-4}}}^{-1/3} \epsilon_e'{^{5/3}}
   A_\nu^{7/12} t_d^{-1/3}.
\end{equation}
Substituting this into equation (\ref{nuca}) we find the cooling frequency in
forward shock for $z=1$:

\begin{equation}
\nu_{cf} = 1.5\times10^{7} {\E_{52}^{1/2}t_d^{7/6}\over
         \epsilon_{B_{f_{-4}}}^{5/6}\epsilon_e'{^{22/3}} A_\nu^{14/3} } 
         \quad{\rm Hz}.
\label{nucf2}
\end{equation}

For $\nu_{cf}<\nu_i$ \& $Y\gg1$, $Y\approx(\epsilon_e/\epsilon_{B_f})^{1/2}$, 
and
\begin{equation}
\nu_{cf} = 2\times10^{9} {\E_{52}^{1/2} t_d^{1/2}\over
      \epsilon_{B_{f_{-4}}}^{1/2}\epsilon_e \epsilon_e'{^4} A_\nu^{7/2} }
      \quad{\rm Hz}.
\label{nucf3}
\end{equation}
For $Y\ll1$ the cooling frequency is obtained by setting $Y=0$ in 
equation (\ref{nuca}).

Since the peak of the $\gamma$-ray spectrum for 021211 is at 
$\nu_{cf}\sim\nu_{if}=1.2\times10^{19}$Hz, we obtain from equation 
(\ref{nucf2}) the following relation
\begin{equation}
\epsilon_e' A_\nu^{7\over11}\epsilon_{B_f}^{5\over44}(t_d) = 
   8.4\times10^{-3} \,\E_{52}^{3\over44} t_d^{7\over44}.
\label{ab1}
\end{equation}
Using equations (\ref{epsilB2}) and (\ref{ab1}), and taking 
$\epsilon_{B_f}(t_d)/ \epsilon_{B_f}(t=11min)\sim 10$
as discussed at the beginning of this subsection, we find
\begin{equation}
 \epsilon_{B_f}^{6\over11}(t_d) \E_{52}^{14\over11} t_d^{7\over11} =
         33\,A_\nu^{6\over11}.
\end{equation}
This equation can be satisfied if we take $\E_{52}=30$,
$\epsilon_{B_f}(t_d)\sim 0.1$, $t_d\sim2$s \& $A_\nu\sim 1$. However,
for these parameters, the density of the medium $A_*\sim 5\times10^{-4}$,
which is three orders of magnitude smaller than the value for a typical
Wolf-Rayet star wind. Thus we find that the $s=2$ model has difficulty 
producing the early optical and $\gamma$-ray flux for 021211.

\subsubsection{Radio flux upper limits}

The FS peak flux for 021211, for $s=2$ medium, declines as $5A_f t^{-1/2}$mJy, 
and therefore the peak flux at 10 day is $\sim5.8\mu$Jy. The expected flux 
at 8.5 GHz at 10 day after the explosion is thus less than 5.8$\mu$Jy, and 
entirely consistent with the upper limit of $\sim 35\, \mu$Jy 
(Fox et al. 2003).

The peak frequency for the RS emission declines as $t^{-17/12}$
and the peak flux declines as $\sim t^{-1}$. Therefore the peak frequency
and flux at 0.1 day are $8 \times 10^8$ Hz and 1.0 mJy, respectively. 
The absorption frequency decreases as $\nu_A\propto t^{-5/6}$ and 
thus $\nu_A\sim 1.1\times10^{11}$ Hz at 0.1 day. Therefore, the
flux at 8.5 GHz at 0.1 day is expected to be about 6$\mu$Jy, 
which is well below the upper limit of Fox \etal (2003).

\subsubsection{A summary of results for $s=2$ medium}

Pre-ejected wind circum-burst medium ($s=2$), for 021211, very easily 
accommodates the flux upper limit in 8.46GHz band reported at 0.1 and 10 days.
However, unlike, $s=0$, it requires extreme parameters -- $\epsilon_{B_f}\sim
1$ \& $A_*\sim 5\times10^{-4}$ -- to explain the $\gamma$-ray \& early
optical radiations. We did not find any common set of parameters, at 11 min,
to explain both the early afterglow and the $\gamma$-ray observations
simultaneously; solutions to explain early optical emission require, 
as in $s=0$, magnetic field energy density in RS to be larger than the 
FS by a factor of a few hundred.

\section{Discussion}

We have shown that for GRB 021211 the gamma-ray burst emission, 
the optical afterglow, and upper limit on radio flux, can be 
understood in terms of emission from forward and reverse
shock regions. To explain the early optical data -- prior to 11
min -- as resulting from the reverse shock, requires a high
energy density in magnetic field, $\epsilon_{B_r}\sim 10^{-1}$, which is
about three orders of magnitude larger than the value we obtain for the
forward shock region at 11 min from optical data. 
Zhang et al. (2003) have suggested that the bright optical flash
in GRB 990123 also requires high magnetic field in reverse shock.

We find that the gamma-ray emission for GRB 021211 can be explained as
synchrotron radiation in the forward shock. The value for 
$\epsilon_B$ in forward shock at deceleration required to explain the
GRB fluence is about 10$^{-2}$ which is larger by about two orders of
magnitude compared with the value at 11 minutes. This suggests that
$\epsilon_B$ in the forward shock is time dependent at early times. The
combination of high $\epsilon_B$ in the ejecta, and high but somewhat 
smaller $\epsilon_B$ in forward shock at deceleration time, suggests that
the magnetic field in the RS is perhaps the frozen-in field of 
the highly magnetized ejecta from the burst as suggested by eg.,
Usov (1992), \Meszaros \& Rees (1997), Lyutikov \& Blandford (2002), and
the field in the early FS could be due to a small mixing of the ejecta
with the shocked ISM.

The transverse magnetic field in the ejecta decays as $R^{-1}$ at early 
times, when the radial width of the material ejected in the explosion
($\delta R$) is constant, and the total energy in magnetic field 
is conserved. Therefore,
if an explosion puts out equal amounts of energy in magnetic field and 
relativistic ejecta, the equipartition will continue to hold until $\delta R$
starts to increase. For a burst of duration $T$ and Lorentz factor $\Gamma$,
$\delta R\sim\max\{cT, R/\Gamma^2\}$, and so $\delta R$ increases only when
$R\gta cT\,\Gamma^2$. At large $R$, the magnetic field decays
as $R^{-2}$ and the energy in magnetic field decreases as $1/R$.
For GRB 021211, $T\sim 4$s \& $\Gamma\sim500$ (see \S6), 
we expect a substantial fraction of the explosion energy 
in magnetic field at the deceleration radius of $\sim10^{17}$cm, if the 
burst was initially poynting flux dominated. 

We note that synchrotron radiation in the reverse shock could also explain
the observed $\gamma$-ray fluence and the spectral peak frequency
for 021211 (but not for the parameter space shown in fig. 2, which has
too little flux at 50 kev). However, the low-energy spectral index in 
this case is $\sim -0.5$, whereas the observed index is 0.4 --- 
$f_\nu\propto\nu^{0.4}$ --- thereby killing this interesting possibility.

 The early and late time optical data is consistent with the density of 
the medium in the vicinity of the bust to be uniform, and a r$^{-2}$
density profile is disallowed; the data can be fitted by a medium
with 1/$r^2$ density profile provided that the density is smaller than
normally expected for Wolf-Rayet stars at a distance of 10$^{17}$cm
by a factor $\gta 10^4$.

The upper limit on radio flux at 8.5 GHz at 10 days (Fox et al., 2003)
poses a problem for the uniform density medium. The solution
requires a combination of jet break at about 10 days and a decrease
in $\epsilon_B$ by a factor of $\sim2$ between 11 min and 10 days, that
we find not particularly appealing.

We can set an upper limit of about 10 on the number of electron \& positron 
per proton in the ejecta. The presence of pairs softens the reverse shock 
synchrotron radiation. When the charged lepton number exceeds about 10, the 
resulting reverse shock emission peaks at too low an energy, and produces
flux at 8.5 GHz at 0.1 day that exceeds the observational upper limit
by more than an order of magnitude.

\acknowledgments{We thank Erin McMahon, Ramesh Narayan and Brad Schaefer for 
useful discussions.}

\vfill\eject
\begin{figure}
\plotone{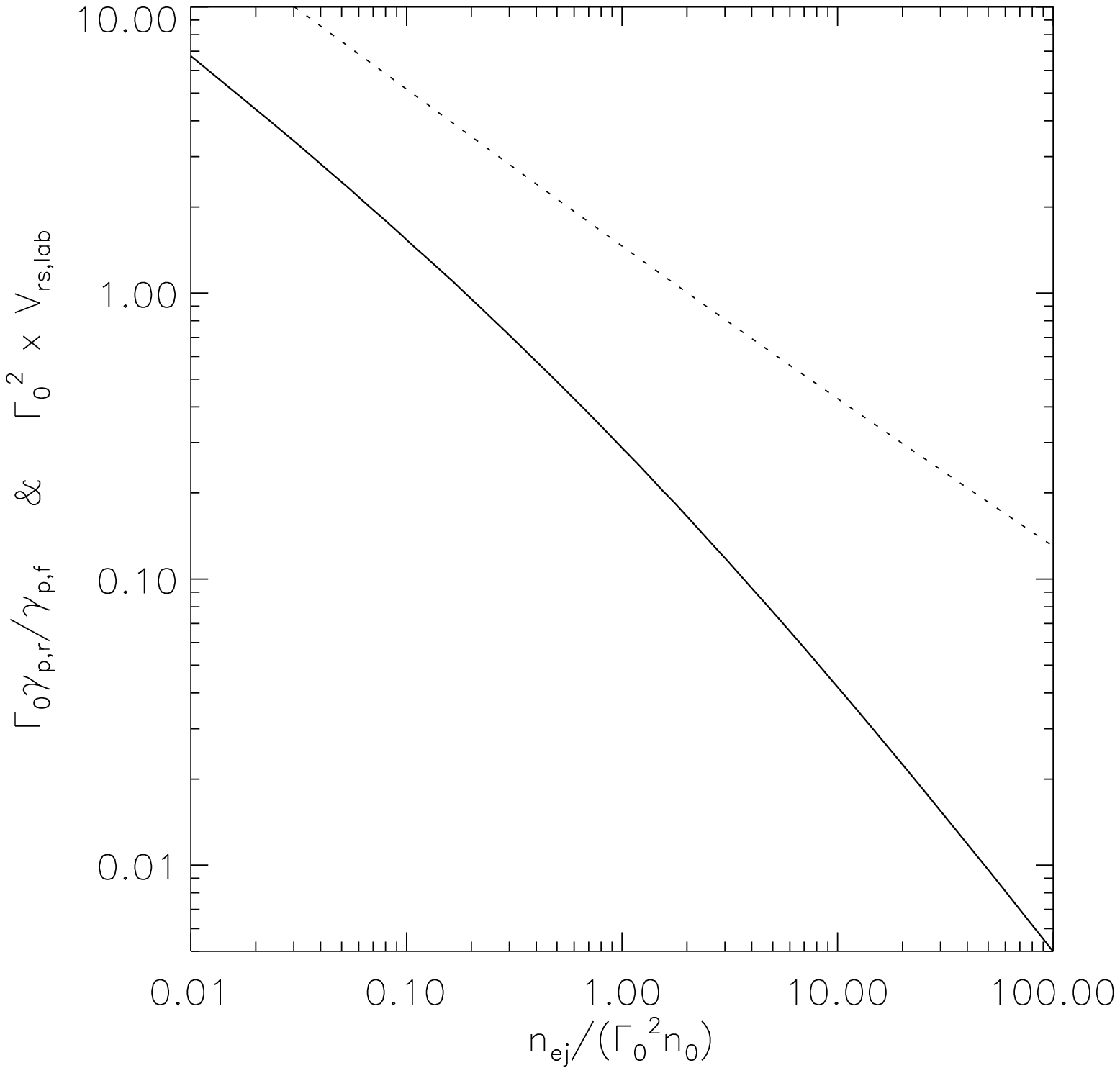}
\caption{\small The solid line is the ratio of the thermal energy per 
proton in the reverse shock and the forward shock,
 $\gamma_{p,r}/\gamma_{p,f}$, as a function of $n_{ej}/n_0$
(the ratio of the comoving frame density of the unshocked ejecta and of 
the circumburst medium). The ejecta initial Lorentz factor, $\Gamma_0$, 
is used to normalize both these ratios such that the curves shown are 
independent of it. To a 
good approximation, $\Gamma_0 \gamma_{p,r}/ \gamma_{p,f} \simeq 0.25 
(n_{ej}/n_0\Gamma_0^2)^{-0.7}$.  The relative velocity of the 
reverse shock front relative to the unshocked ejecta, as measured in the 
lab frame, $V_{rs,lab}$, is shown by the dotted line;  
$\Gamma_0^2 V_{rs,lab} \simeq 1.4 (n_{ej}/\Gamma_0^2 n_0)^{-0.5}$. 
}
\end{figure}

\begin{figure}[h]
\vbox to2.6in{\rule{0pt}{2.6in}}
\includegraphics{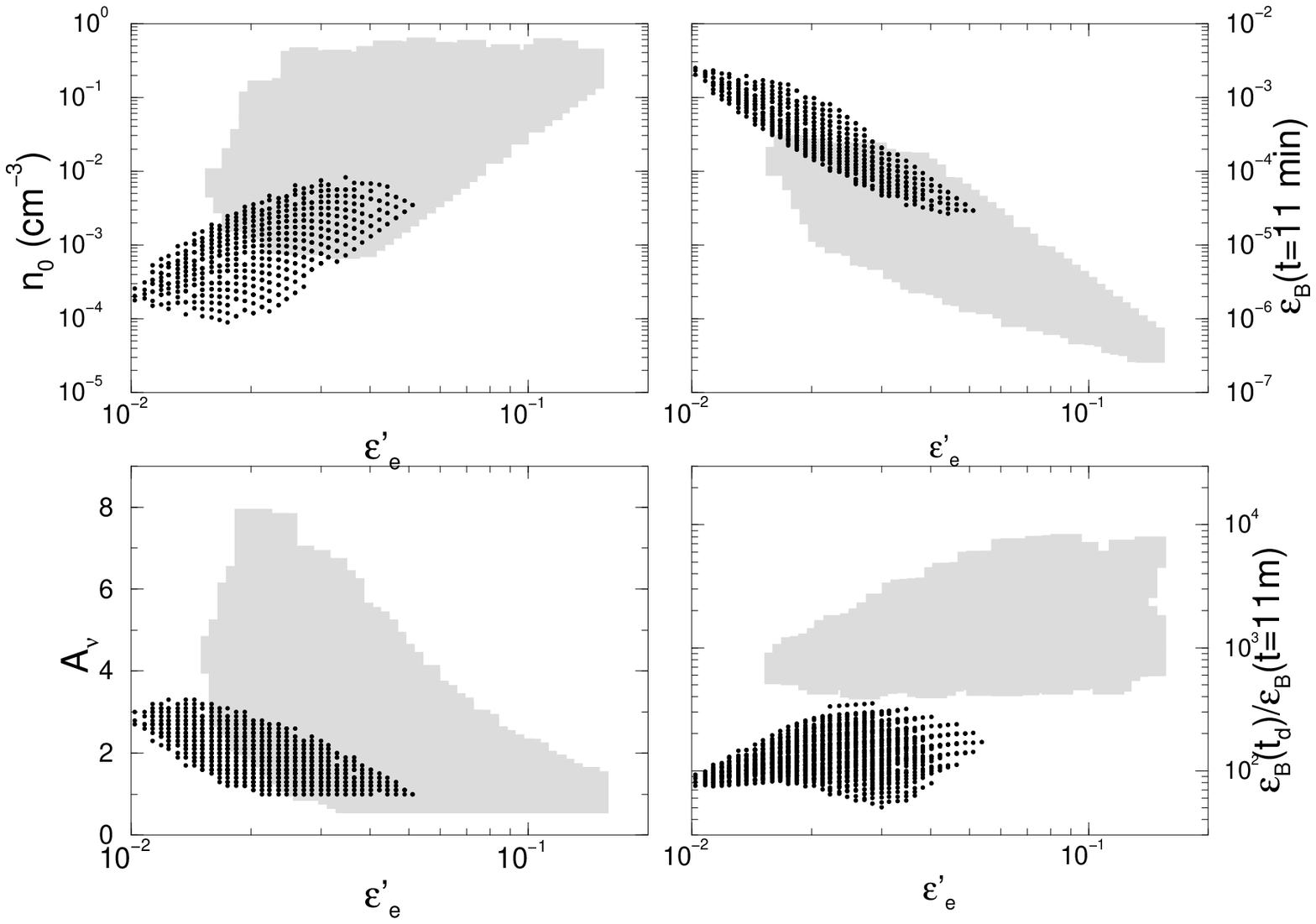}
\vspace*{34mm}
\figcaption{\small Grey area: parameter space (and derived quantities) 
for a {\it homogeneous} external medium allowed by the observed R-band 
flux from reverse shock (RS) at 90 seconds and from forward shock (FS) 
at 11 minutes, but not including the $\gamma$-ray flux during the burst.
Black dots show the parameter space allowed for the observed gamma-ray 
flux to arise in forward shock as synchrotron emission; inverse Compton 
emission in FS or RS cannot account for the observed $\gamma$-ray spectrum.
Top left panel shows the allowed density for the ISM ($n_0$) and $\epsilon'_e$
(which gives the minimum LF of shock heated electrons); gray area is allowed
by the early reverse shock, and late forward shock, optical observations,
and black dots show the region of the parameter space permitted by GRB
observation. The top right panel shows the allowed value for the
magnetic field parameter $\epsilon_B$ in the forward shock at 11 minutes.
The lower left panel shows $A_\nu$, a parameter that specifies the
peak of the synchrotron spectrum at 11 minutes (the peak frequency is
4.95x10$^{14}/A_\nu$ Hz). The lower right panel shows the ratio of
$\epsilon_B$ in RS and FS at 11 minutes (grey region), and the
ratio of $\epsilon_B$ in FS at deceleration time (~3s) and at 11 minutes.
Note that $\epsilon_B$ in RS is larger by a factor of about 10$^3$ compared
with the value in FS at 11 minutes, and $\epsilon_B$ in FS at
deceleration is larger than at 11 minutes by a factor of about 10$^2$.
$\E_{52}=30$, $t_d=3$s, $p=2.5$ \& $z=1.0$ for all calculations; $p=2.2$
gave very similar results.
}
\end{figure}


\begin{references}

\reference{} Crew, G. \etal 2003, ApJ, submitted (astro-ph/0303470) 
\reference{} Fox, D. \etal 2003, ApJ, 586, L5
\reference{} Fruchter, A. \etal 2002, GCN 1781
\reference{} Greiner, J. \etal 2003, GCN 2020
reference{} Kobayashi, S. 2000, ApJ 545, 807
\reference{} Li, W., Filippenko, A., Chornock, R. \& Jha, S. 2003, ApJ, 
             submitted (astro-ph/0302136)
\reference{} Lyutikov, M. \& Blandford, R.D. 2002, in ``Beaming and Jets in
  Gamma-ray Bursts'', R. Ouyed, J. Hjorth \& A. Nordlund, eds. astro-ph/0210671
\reference{} Matheson, T. \etal 2003, GCN 2120
\reference{} \Meszaros, P. 2002, ARA\&A 40, 137
\reference{} \Meszaros, P. \& Rees, M.J. 1993, ApJ 405, 278
\reference{} \Meszaros, P. \& Rees, M.J. 1997, ApJ 482, 29
\reference{} Panaitescu, A. \& Kumar, P. 2002, ApJ 571, 779
\reference{} Panaitescu, A. \& \Meszaros, P., 1998, ApJ 492, 683
\reference{} Panaitescu, A. \& \Meszaros, P., 2000, ApJ 544, L17
\reference{} Piran, T. 1999, Physics Reports, 314, 575
\reference{} Rees, M.J. \& \Meszaros, P. 1994, ApJ 430, L93
\reference{} Sari, R., \& Piran, T. 1999, ApJ 517, L109
\reference{} Usov, V.V. 1992, Nature 357, 472
\reference{} Wijers, R., Rees, M.J., \& \Meszaros, P. 1997, MNRAS, 288, L51 
\reference{} Zhang, B., Kobayashi, S. \& \Meszaros, P., 2003, astro-ph/0302525
\reference{} Wozniak, P. \etal 2002, GCN 1757  

\end{references}
\end{document}